\documentstyle[12pt,aaspp4,epsfig]{article}

\newcommand{\etal}{{\it{et al.}}~}

\renewcommand{\sun}{\odot}
\newcommand{\Msun}{\mbox{$M_\sun$}}

\def\gtsima{$\, \buildrel > \over \sim \,$}
\def\ltsima{$\, \buildrel < \over \sim \,$}
\def\simgt{\lower.5ex\hbox{\gtsima}}
\def\simlt{\lower.5ex\hbox{\ltsima}}

\begin{document}

\title{Observations and Implications of the Star Formation History of the LMC
\footnote{Based on observations with the NASA/ESA {\it Hubble Space Telescope},
obtained at the Space Telescope Science Institute, operated by AURA Inc under
contract to NASA}}
\author{
Jon~A.~Holtzman\altaffilmark{2},
John~S.~Gallagher~III\altaffilmark{3},
Andrew~A.~Cole\altaffilmark{3},\\
Jeremy~R.~Mould\altaffilmark{4},
Carl~J.~Grillmair\altaffilmark{5},
Gilda~E.~Ballester\altaffilmark{6},
Christopher~J.~Burrows\altaffilmark{7},\\
John~T.~Clarke\altaffilmark{6},
David~Crisp\altaffilmark{8},
Robin~W.~Evans\altaffilmark{8},
Richard~E.~Griffiths\altaffilmark{9},\\
J.~Jeff~Hester\altaffilmark{10},
John~G.~Hoessel\altaffilmark{3},
Paul~A.~Scowen\altaffilmark{10},\\
Karl~R.~Stapelfeldt\altaffilmark{8},
John~T.~Trauger\altaffilmark{8},
and Alan~M.~Watson\altaffilmark{11}
}
 
\altaffiltext{2}{Department of Astronomy, New Mexico State
University, Dept 4500
 Box 30001, Las Cruces, NM 88003, holtz@nmsu.edu}
\altaffiltext{3}{Department of Astronomy, University of Wisconsin
-- Madison, 475 N. Charter St., Madison, WI 53706,
jsg@tiger.astro.wisc.edu,cole@ninkasi.astro.wisc.edu,hoessel@jth.
astro.wisc.edu}
\altaffiltext{4}{Mount Stromlo and Siding Spring Observatories,
Australian National University, Private Bag, Weston Creek Post
Office, ACT 2611, Australia,
jrm@mso.anu.edu.au}
\altaffiltext{5}{SIRTF Science Center, Caltech, MS 100-22, Pasadena, CA 91125,
carl@ipac.caltech.edu}
\altaffiltext{6}{Department of Atmospheric, Oceanic, and Space
Sciences,
University of Michigan, 2455 Hayward, Ann Arbor, MI 48109,
gilda@sunshine.sprl.umich.edu, clarke@sunshine.sprl.umich.edu}
\altaffiltext{7}{Astrophysics Division, Space Science Department,
ESA \& Space
Telescope Science Institute, 3700 San Martin Drive, Baltimore, MD
21218,
burrows@stsci.edu}
\altaffiltext{8}{Jet Propulsion Laboratory, 4800 Oak Grove Drive,
Pasadena,
CA 91109, carl@wfpc2-mail.jpl.nasa.gov, dc@cov,
rwe@wfpc2-mail.jpl.nasa.gov, krs@wfpc2-mail.jpl.nasa.gov,
jtt@wfpc2-mail.jpl.nasa.gov}
\altaffiltext{9}{Department of Physics, Carnegie Mellon
University, 5000 Forbes
Ave, Pittsburgh, PA 15213}
\altaffiltext{10}{Department of Physics and Astronomy, Arizona
State University,
Tyler Mall, Tempe, AZ 85287, jjh@cosmos.la.asu.edu,
scowen@tycho.la.asu.edu}
\altaffiltext{11}{Instituto de Astronom\'\i a UNAM, J. J. Tablada
1006, Col. Lomas de Santa Maria, 58090 Morelia, Michoac\'an,
Mexico, alan@astrosmo.unam.mx}

\newpage
\begin{abstract}

We present derivations of star formation histories based on
color-magnitude diagrams of three fields in the LMC from HST/WFPC2
observations. One field is located in the LMC bar and the other two are
in the outer disk. We find that a significant component of stars older
than 4 Gyr is required to match the observed color-magnitude diagrams.
Models with a dispersion-free
age-metallicity relation are unable to reproduce the width of the
observed main sequence; models with a range of metallicity at a given
age provide a much better fit. Such models allow us to construct
complete ``population boxes'' for the LMC based entirely on
color-magnitude diagrams; remarkably, these qualitatively reproduce the
age-metallicity relation observed in LMC clusters.  We discuss some of
the uncertainties in deriving star formation histories by our method
and suggest that improvements and confidence in the method will be
obtained by independent metallicity determinations.  We find,
independently of the models, that the LMC bar field has a larger
relative component of older stars than the outer fields. The main
implications suggested by this study are: 1) the star formation history
of field stars appears to differ from the age distribution of clusters,
2) there is no obvious evidence for bursty star formation, but our
ability to measure bursts shorter in duration than $\sim$ 25\% of any
given age is limited by the statistics of the observed number of stars,
3) there may be some correlation of the star formation rate with the last 
close passage of the LMC/SMC/Milky Way, but there is no dramatic effect,
and 4) the derived star formation
history is probably consistent with observed abundances, based on
recent chemical evolution models.

\end{abstract}

\section{INTRODUCTION}

Recent improvements in data quality and analysis tools have opened up
the possibility of deriving detailed star formation histories for Local
Group galaxies based on observed colors and brightnesses of individual
stars. Results have been somewhat surprising, indicating a wide 
diversity of star formation histories among galaxies of the Local
Group, even for galaxies of a given morphological type (e.g., the dwarf
spheroidals); some recent reviews have been presented by Mateo (1998),
Grebel (1998), and Da Costa (1998).

The Large Magellanic Cloud occupies a special role in these studies.
As our nearest neighbor (apart from the Sagittarius dwarf), it allows
observations of faint stars which include essentially unevolved stars
that are fainter than the turnoff corresponding to the age of the
universe. Such stars contain information about the initial mass
function and also about the earliest stages of the star formation
history of a galaxy. In addition, we can obtain accurate photometry of
stars down to the oldest main sequence turnoff.  It is critical that
information derived from these stars agrees with star formation
histories derived from brighter, more evolved stars if we are to
believe the results on star formation histories derived for more
distant galaxies, where only the brighter stars are observable. This is
especially true given that uncertainties in our understanding of
stellar evolution are generally larger for stars in their later stages
of evolution, when they are brighter.

In addition, the LMC provides a unique opportunity to compare the star
formation history of its field population with that of its star
clusters, since the LMC has a rich population of the latter. This has
implications for understanding whether clusters form in a different mode
of star formation than field stars, and is important to understand the
degree to which one can trace the global star formation history of a
galaxy from its constituent star clusters.

Several studies have suggested star formation histories for the
field population in the LMC. Early studies by Butcher (1977) and Stryker
(1984) suggested that the LMC might be composed primarily of younger
stars.  Using ground-based observations which did
not quite reach to the oldest main sequence turnoff, Bertelli et al.
(1992) and Vallenari et al. (1996a,b) suggested that the LMC field
population was composed primarily of younger stars with ages less than
a typical burst age of 4 Gyr, with some indication that the burst age
varied across the galaxy. Deeper observations with the HST have not
confirmed this picture, instead suggesting that star formation has been
more continuous over the lifetime of the LMC, although almost certainly
with an increase in star formation rate in the past several Gyr
(Holtzman et al. 1997, Geha et al. 1998). All of these studies were for
regions outside the LMC bar. Inside the bar, Olsen (1999) suggests that
the star formation rate has also been more continuous, possibly
extending back for a longer period at a roughly constant rate than the
outer fields; this differs from the conclusions of Elson et al (1997),
who suggest that the bar formed relatively recently and has an age of
$\sim$ 1 Gyr.

In this paper, we present derivation of the star formation history for
a field in the LMC bar observed with the HST/WFPC2, as well as for
several previously published outer fields, using a more detailed analysis of
the color-magnitude diagram. Along with this, we discuss some of the
problems and limitations of the techniques (including
ours) which are being used to extract star formation histories.  We
attempt to present a summary of some of the main implications of recent
results on the star formation history of the LMC:  the relation between
the field and cluster star formation history, differences between the
outer regions and the bar, the relationship of the star formation
history to the dynamical history of the LMC, and the relation between
the star formation history derived from studies of color-magnitude
diagrams with that derived from chemical evolution studies.

\section{OBSERVATIONS}

All of the data discussed in this paper were obtained with the Wide
Field/Planetary Camera 2 (WFPC2) on the Hubble Space Telescope (HST)
as part of Guaranteed Time Observations granted to the WFPC2
Investigation Definition Team. Three separate fields were observed,
with two located several degrees from the LMC center, and one located
within the LMC bar; details are given in Table 1 (for a map of the locations 
of the outer fields, see Geha et al. 1998)  In all fields,
observations were made through the F555W and F814W filters.  Standard
reduction procedures were applied to all of the frames, as discussed by
Holtzman et al (1995a).

The two outer fields are relatively uncrowded. In contrast, the bar
field is fairly crowded, so stars cannot be seen as faint as in the
outer field. This problem is exacerbated by the fact the the PSF in
the bar exposures is significantly broader than in the other fields.
This presumably occurred because of a large focus excursion (the 
so-called ``breathing'' of the HST secondary); to our knowledge, these
frames represent one of the largest examples of this behavior. These
exposures serve as a distinct warning to those who assume that the
HST PSF is temporally stable.

\section{ANALYSIS}

\subsection{Photometry}

Photometry in each of the fields was done using profile-fitting photometry
as described by Holtzman et al. (1997). To summarize, we performed
the photometry simultaneously on the entire stack of frames taken in each
field, solving for the brightnesses in the two colors, the relative positions
of the stars, and the frame-to-frame positional shifts (and scale changes
between the filters), using an individual custom model PSF for each frame.
The model PSFs used a separate focus for each frame as derived using
phase retrieval of a few stars in the frame; the models also
incorporate the field dependence of the pupil function as specified by
the WFPC2 optical prescription, and the field dependence of aberrations
as derived from phase retrieval from some other stellar fields. Instrumental
magnitudes were placed on the WFPC2 standard system using the
calibration of Holtzman et al. (1995b). No correction was made for
the possible effect of charge transfer efficiency problems, since these
are expected to be relatively small for the background levels in our frames,
especially for the relatively bright stars on which most of our analysis
is based.  The software for all of the PSF modelling and photometry 
was implemented in the XVISTA image processing package.

\begin{figure}
\caption{Observed color magnitude diagram for combination of the two outer
fields (left) and the bar field (right).}
\plotone{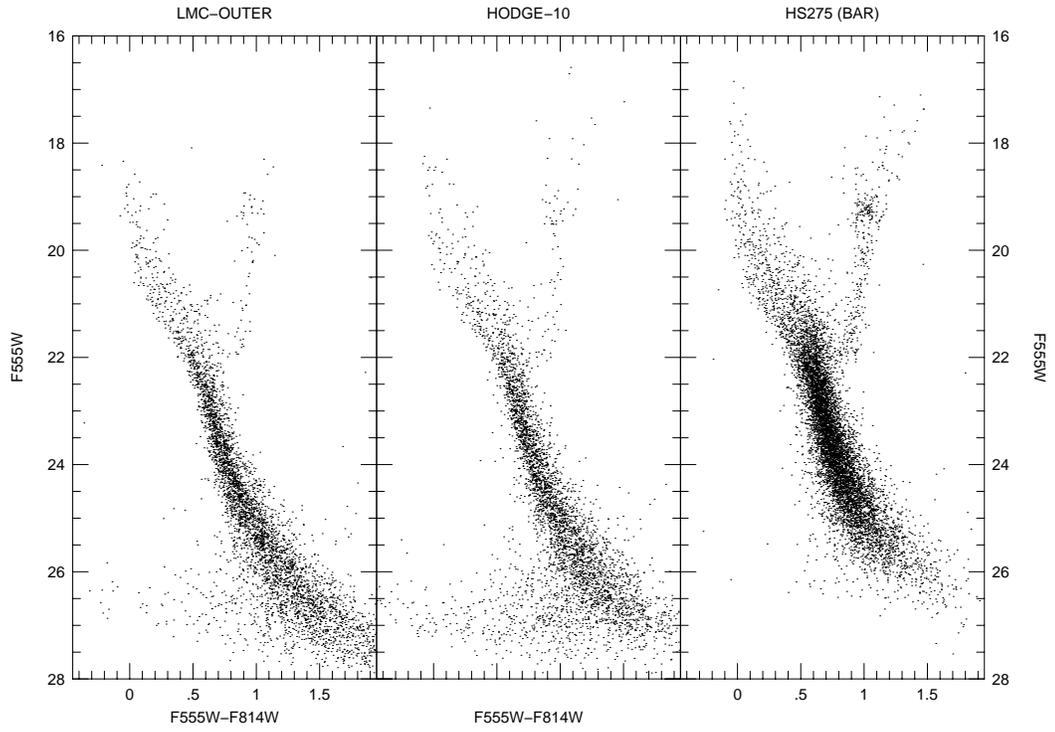}
\label{fig-cmd}
\end{figure}

Figure \ref{fig-cmd} shows the derived color-magnitude diagrams for the 
three fields.

We investigated errors in the photometry and its completeness using a
series of artificial star tests, in which artificial stars were placed
into each image at a range of brightnesses and the photometry was
redone. The individual errors for all of the artificial stars (observed
vs. input brightness) were tabulated for use in the construction of
artificial color-magnitude diagrams, as discussed below.

\subsection{Derivation of star formation histories}

Various groups in the past several years have published descriptions of
techniques used to infer star formation histories based on the
distribution of stars in a color-magnitude diagram, or, more generally,
for observations of stars in multiple colors (e.g., Tolstoy and Saha 1996;
Dolphin 1997; Hurley-Keller, Mateo, \& Nemec 1998; Ng 1998; 
Olsen 1999; Gallart et al. 1999; Han, in prep.). 
These are all similar in concept, in that a set of observations
is fit using some combination of individual simple stellar populations in
an effort to derive the relative importance of different simple stellar
populations and thus a star formation history.  The techniques differ
in detail, using different metrics against which one measures how well
a given model matches the observations (e.g., maximum likelihood,
minimum $\chi^2$ for different bins in color-magnitude space, etc.), different
techniques with which the best fit is sought (e.g., linear least
squares, genetic algorithm, trial and error), and different input
stellar models.

For several years, we have also been doing fits for star formation
histories.  Our technique bins stars in a color-magnitude diagram, and
searches for a best fit by minimizing the $\chi^2$ between the number
of observed and model stars in the different bins. The search for the
best fit is automated, using nonlinear least squares to solve for the
the relative amplitudes of each simple stellar population, and optionally
for the distance, reddening, and metallicity; the fit is nonlinear because
the problem is formulated to insure that only positive amplitudes for
each simple stellar population are allowed.  

For our input stellar models, we use the isochrones from the Padova
group (Bertelli et al.  1994), and recently have also experimented with
a preliminary version of the newest isochrones from this group
(Girardi, private communication). We find that the isochrones from
Girardi provide a significantly better match to the observed giant
branches, as they predict hotter temperatures for the giants; around the
main sequence, where most of our stars are located, the Girardi isochrones
are similar to the original Padova set.
In all of our subsequent analysis, we use the Girardi isochrones which
were available to us (Z=0.001, 0.004, and 0.008) in conjunction with
Padova isochrones for other metallicities (Z=0.0004, 0.02, 0.05).
The isochrones are used in conjunction with Kurucz (1993) model
atmospheres to derive colors and brightnesses in the WFPC2 photometric
system.  

We allow for arbitrary ages and metallicities by interpolating
within the isochrones using a set of equivalent evolutionary points to
preserve the correct isochrone shape. Our basis simple stellar
populations are either discrete bursts or are constructed assuming
continuous star formation within specified epochs.  Typically, we use
epochs with equal widths in the logarithm of the age, to account for
the fact that isochrones at a fixed age difference become more similar
as a population ages. For the current work, we have assumed a Salpeter
initial mass function ($dN/dM \propto M^\alpha$, with $\alpha=-2.35$)
for most of our models, although we have tried
some other IMFs as well, as discussed briefly below. Uncertainty about
the IMF is probably responsible for one of the largest sources of
potential errors in our results.

Given model predictions for the number of stars as a function of color
and magnitude for any given star formation history, along with a
distance and extinction, we account for observational errors and
incompleteness by smearing the model results with the errors derived
from our artificial star tests.  We use the exact tabulations of
measured errors for our artificial stars to do this smearing, and thus
make no assumption that the errors are distributed normally (which they
are often not, in particular, because of errors from crowding which are
correlated for each color). The fraction of detected artificial stars
is included in the smearing, so incompleteness is automatically taken
into account.

Previous usages of this software (Holtzman et al. 1997, Geha et al. 1998)
have used it in a mode where the bins are very wide in color, effectively
making this perform a fit to the luminosity function. Fits to the 
luminosity function are less sensitive to possible errors in reddening,
photometric calibration, and color errors in the stellar models. Of
course, throwing away color information lowers the ability to discriminate
among different models. Fits to the full color-magnitude diagram using
narrower color bins, along with a discussion of possible problems with 
interpreting these, are presented next.

\section{DISCUSSION}

\subsection{Derivation of star formation histories from model fits}

We derived two separate star formation histories, one for the ``outer''
fields and one for the bar. For the outer fields, we combined the data
from the two observed fields because there is no strong evidence that
the star formation history of these fields differs (Geha et al. 1998)
and because the relatively smaller number of stars in these fields
limits the accuracy of the derived star formation histories.

It is possible to attempt to determine the distance modulus and/or
the extinction by allowing these to be free parameters in the fit or
by comparing the residuals for various choices of these parameters.
We do this below in an effort to estimate some of uncertainties in
our results. However, we believe that it is important, when possible,
to use additional constraints on these quantities rather than simply
allowing them to be free parameters. There are many methods for
estimate distances and/or extinctions which do not depend on interpretation
of a color-magnitude diagram, for example, the use of variable stars
or HI column densities. As we will argue below, it seems prudent to use 
independent information when possible to constrain the star formation
histories.

For the LMC, however, there is significant debate about the distance
and some uncertainty about the extinction. For our initial fits, we
decided to fix the distance to the LMC at a (extinction-free) distance
modulus of 18.5 and initially chose an extinction of E(B-V)=0.10 based
on the maps of Schwering and Israel (1991) which suggest a foreground
reddening of E(B-V)=0.07; we included an additional 0.03 mag as an
estimate for internal reddening.  However, with these choices we found
that the model main sequences were too red, which, as discussed below,
affected the derived star formation histories in systematic ways. The
only way we could match the color of the main sequence was to adopt
lower extinctions; we adopted E(B-V)=0.04 for the outer fields and E(B-V)=0.07
for the bar field.  Interestingly, ground-based studies of the LMC have
been yielding relatively low reddenings for fields away from young
associations (Zaritsky, private communication), although we do not have
any direct estimates from these studies for our fields at this time.  A
possible alternative to errors in our assumed reddening is errors in
our photometric zeropoints; in fact, a small color change of $\sim$
0.02 mag in F555W-F814W relative to our adopted calibration is
suggested by Stetson (1998). In any case, we find that our derived
results are relatively insensitive to the source of the problem, and
have chosen to use the empirical lower reddenings. 

In a subsequent section, we will present star formation histories
derived from a variety of choices of distance and extinction, in an
effort to understand the uncertainties in our results. Here, we note
that our adopted choices for these quantities actually are those for which
the best-fitting star formation histories give relatively low $\chi^2$ 
values; there is some variation in which distance and extinction give the
absolutely lowest $\chi^2$ depending the field being fit and assumptions
about the age-metallicity relation.

For our initial fits, we constrained the basis stellar populations to
have the age-metallicity relation presented by Pagel and Tautvaisiene 
(1998), who derived it from a chemical evolution model designed to to
fit a variety of observations of LMC clusters and field stars. For our
initial models, we assumed that the age-metallicity relation is
dispersion-free; the cluster data (e.g., Olszewski 1993), however,
suggest that there may be a range of metallicities at any given age.
To obtain isochrones at the appropriate ages and metallicities, we
interpolated among the Padova isochrones. We used age bins with a width
of 0.1 in $\log(age)$, assuming constant star formation within each age
bin.  We binned the data in the color-magnitude diagrams with bin
widths of 0.04 mag in color and 0.06 mag in brightness; these were
compromise values given the observed number of stars.

\begin{figure}
\caption{Derived star formation history for bar field, assuming an 
age-metallicity relation for the LMC. See text for description of the 
various panels.}
\plotone{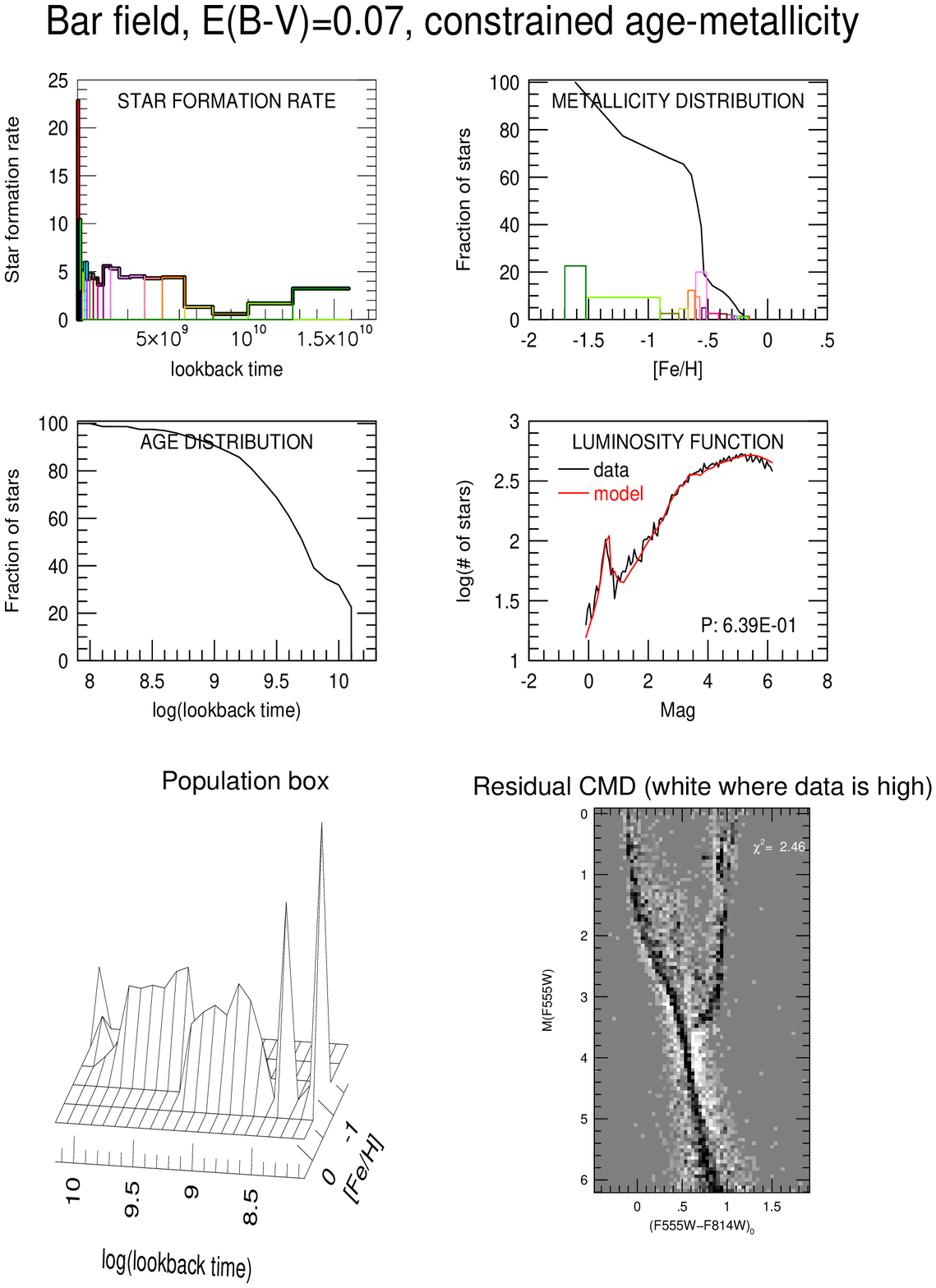}
\label{fig-hs2750_44}
\end{figure}

\begin{figure}
\caption{Same as Fig. 1, but for the outer fields.}
\plotone{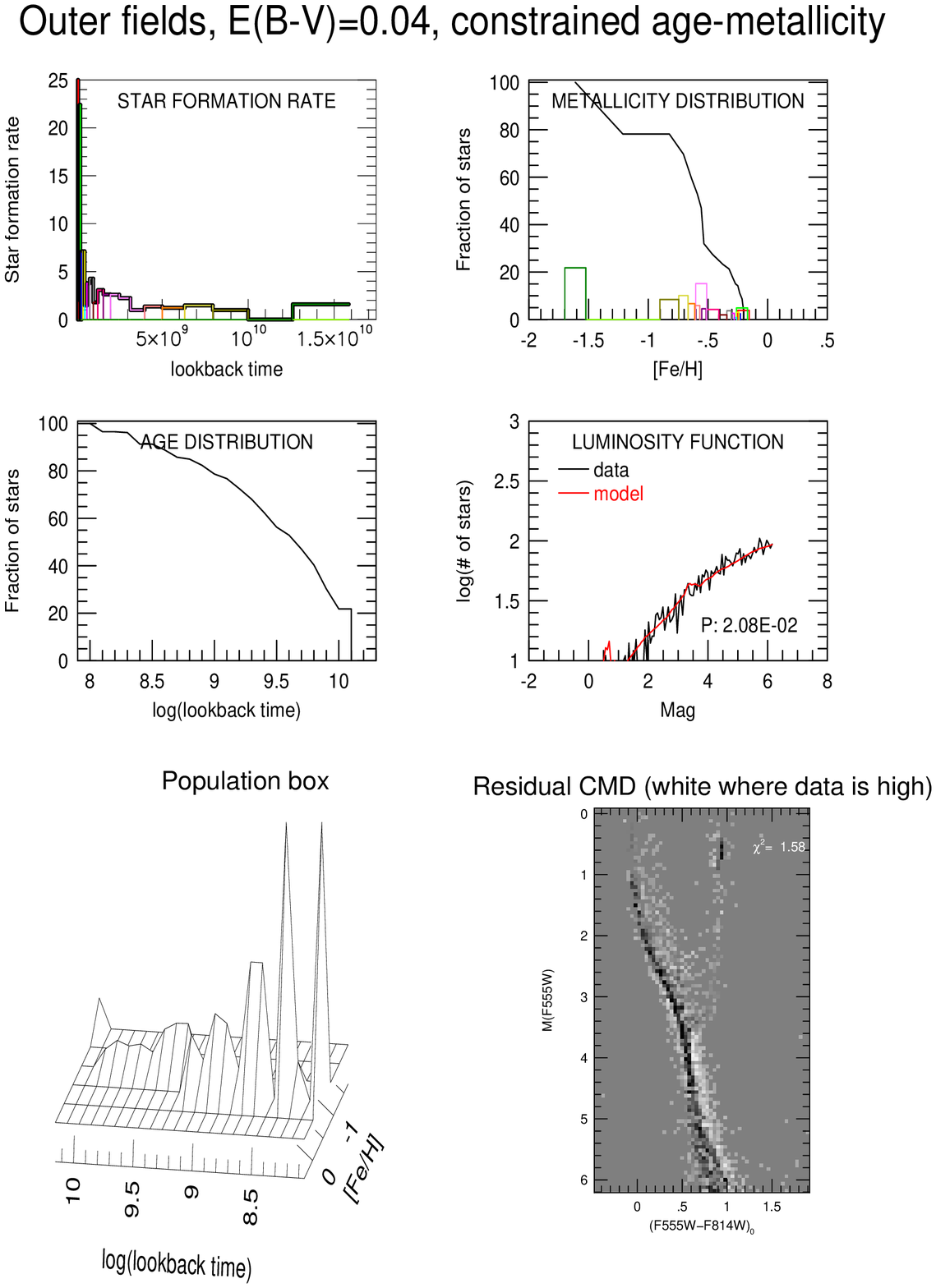}
\label{fig-outer0_47}
\end{figure}

Figures \ref{fig-hs2750_44} and \ref{fig-outer0_47} show a summary of
the derived star formation history information for the two fields with
the constrained age-metallicity relation.  The upper left plot shows
the derived star formation rate as a function of time (abscissa is
linear in lookback time), the lower left shows the cumulative number of
stars formed as a function of log(lookback time), the upper right plot shows
the differential and cumulative metallicity distribution functions, and
the lower right shows the observed and model F555W luminosity functions
(corrected for reddening). The
3D diagram at the lower left shows the ``population box'' (Hodge 1989)
which gives the star formation rate as a function of log(lookback time)
and metallicity; older ages are at the left side of the plot, and lower
metallicities are at the back.  The greyscale  at the lower right shows
the difference between the model and the observed Hess diagrams divided
by the square root of the data. Thus, the greyscale diagram gives the
deviation between the data and the model in units of the deviation
expected from counting statistics. The greyscale is in the sense that
bright areas are regions where the observed number of stars is larger
than the model; the full range from black to white is $-3\sigma$ to
$+3\sigma$.  The quality of the fits is estimated by a reduced $\chi^2$
statistic which is shown in the greyscale diagram, and also by the
probability (given in the luminosity function panel) that the observed
and the model luminosity functions are drawn from the same population,
as inferred from a Kolmogorov-Smirnov test.

These model fits suggest that star formation has been ongoing in the
LMC over its entire history, with fluctuations of a factor of a few in
star formation rate and a higher star formation rate in the past few
Gyr. In the outer fields, there is evidence for an increasing star
formation rate over the past few Gyr, whereas in the bar, the star
formation rate seems to have been more constant recently. More
generally, these fits suggest, as did previous studies (Holtzman et al.
1997, Geha et al. 1998, Olsen 1999), that a significant fraction
($\sim$ 50\%) of the stars in the LMC are relatively old, i.e. older
than $\sim$ 4 Gyr.

However, if one inspects the residual greyscale Hess diagrams, one one
can clearly see some systematic problems with the model fits. In
particular, the model main sequences are much narrower than the observed
main sequences. A similar effect, although at a reduced level, can be
seen in the residuals shown in Olsen (1999); the apparently smaller
effect in those data is plausibly explained by the fact that their
exposure times were shorter (by a factor of $\sim$ 4), leading to
larger photometric errors and hence broader observed and model
sequences.

There are several possible explanations:
\begin{enumerate}
\item The LMC has stars with a \textit{range} of metallicities at any 
given age.
\item A significant number of stars in the LMC are unresolved binary
stars.
\item There is a spread in distances and/or extinctions. This is
unlikely given the inclination of the LMC and the relatively low
total extinction towards our fields.
\item Our observational errors have been significantly underestimated;
we believe this is unlikely given our careful analysis of photometric
errors.
\end{enumerate}

We discuss the first two possibilities in the next sections.

\subsubsection{Metallicity dispersion}

We feel that the mostly likely source of the broad main sequence is that the
LMC has stars with a \textit{range} of metallicities at any given age.
Certainly, the Milky Way has a very significant dispersion in its
age-metallicity relation.  To test whether a spread in metallicities
can account for the observed broad main sequence, we performed the fits
allowing for multiple combinations of age and metallicity. We used the
same age bins as before, but at each age, allowed populations with
discrete metallicities of Z=0.0004, 0.001, 0.004, and 0.008 , 0.02, and 0.05.
The choice of these six metallicities was motivated by the fact that these
were available without any interpolation in metallicity. We allowed for 
all combinations of age with these metallicities.

\begin{figure}
\caption{Derived star formation history for bar field, allowing for multiple
combinations of age and metallicity.}
\plotone{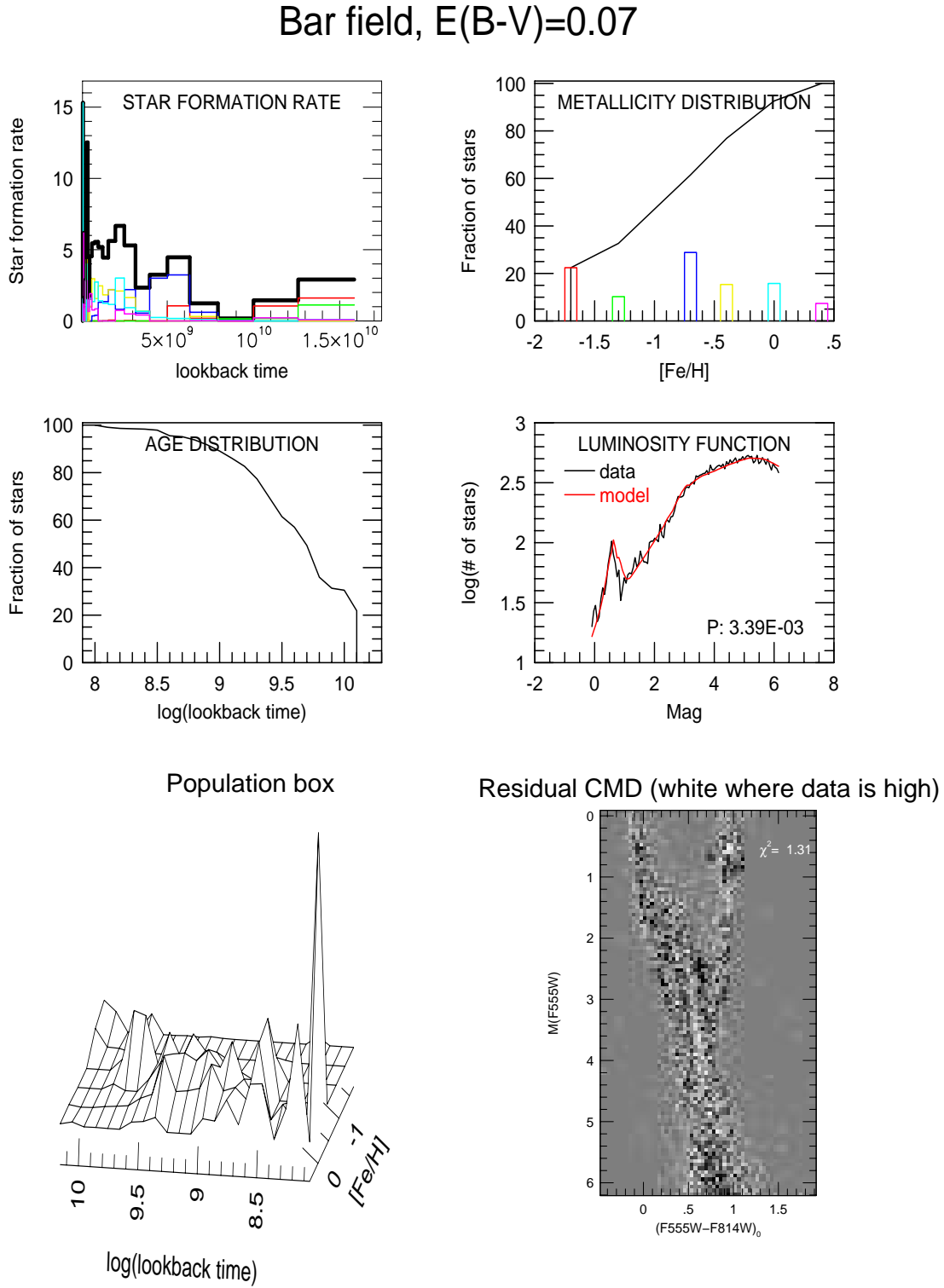}
\label{fig-hs2750_54}
\end{figure}

\begin{figure}
\caption{Same as Fig. 3, but for the outer fields.}
\plotone{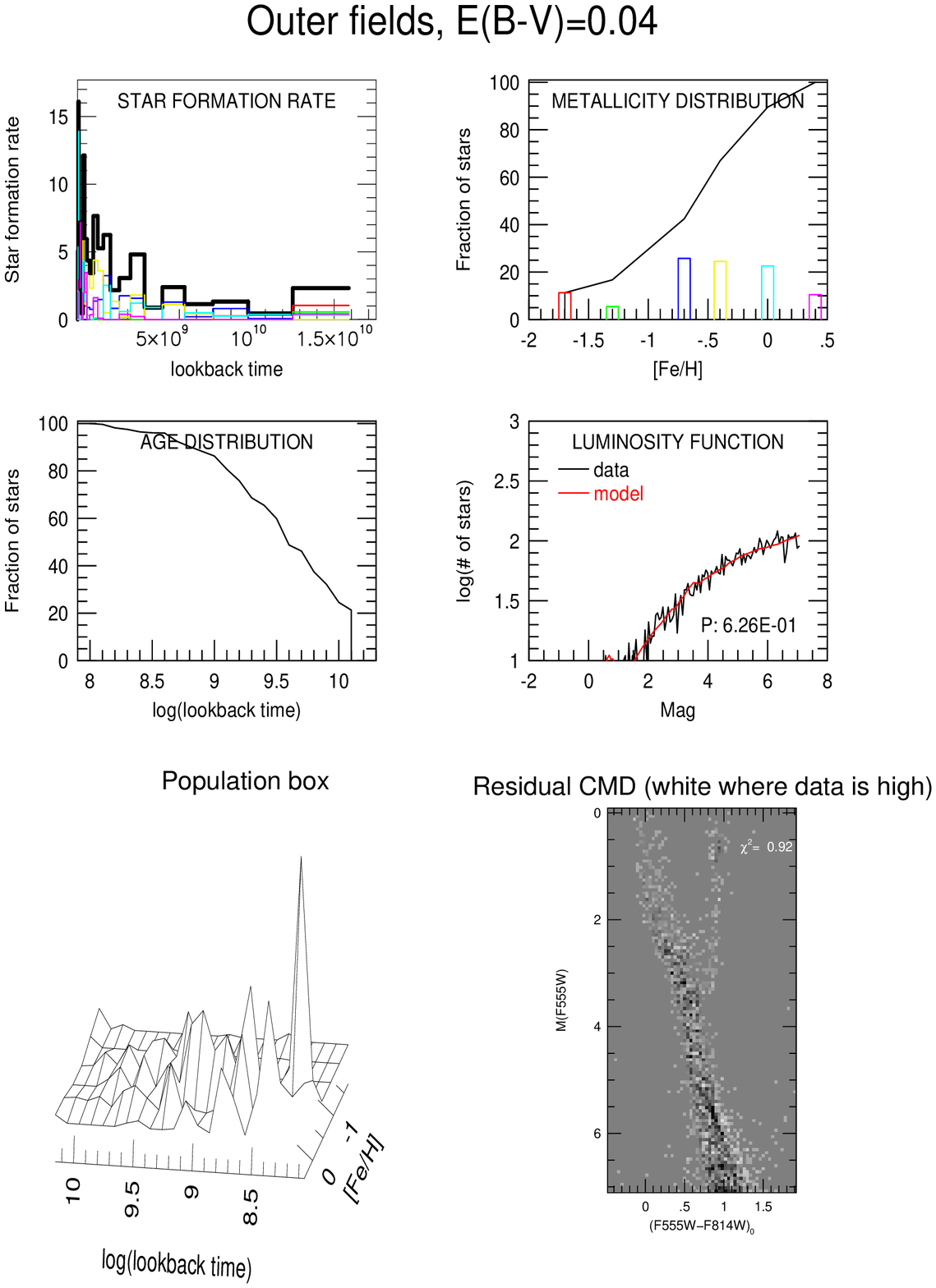}
\label{fig-outer0_57}
\end{figure}

\begin{figure}
\caption{Our derived age-metallicity relation for the bar field. Squares
represent the peak metallicity seen at each age, while crosses represent
the mean metallicity. The solid line gives the relation of Pagel and
Tautvaisiene (1998).}
\plotone{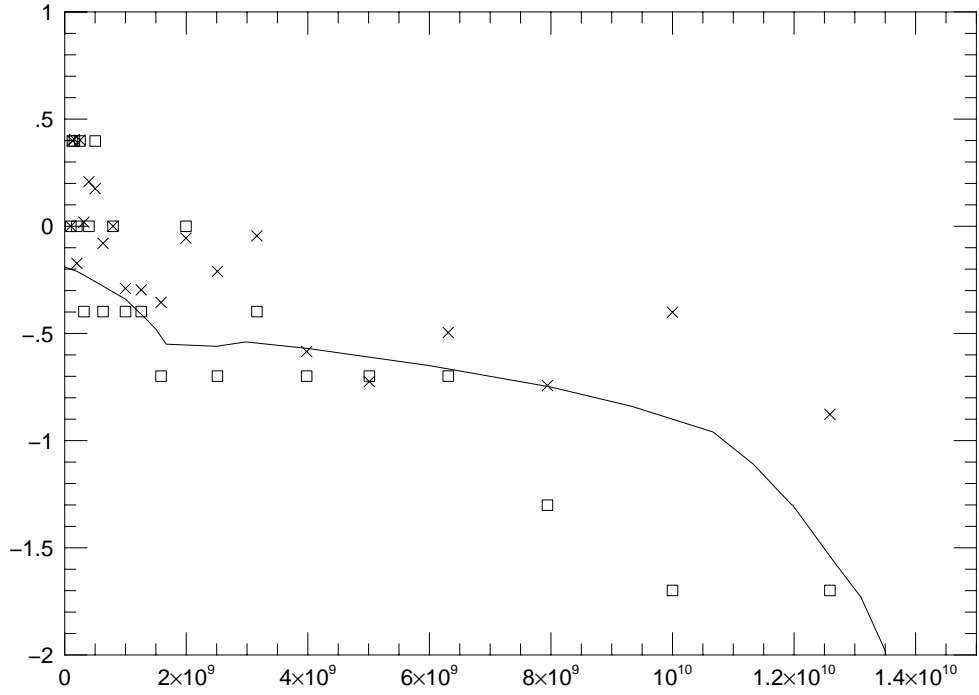}
\label{fig-z}
\end{figure}

Figures \ref{fig-hs2750_54} and \ref{fig-outer0_57} show the results
using these models. Similar star formation histories are derived, but
the resulting residuals show substantially smaller systematic
deviations. Remarkably, the model fits qualitatively recover the mean
age-metallicity relation observed in LMC clusters, despite the fact
that no assumptions were made about this relation at all. This is
demonstrated by comparing the ``population boxes'' for these models
with those using the constrained age-metallicity relation. Although
the derived relation is not especially quantitative since we
only included six discrete metallicities, Figure \ref{fig-z} shows a 
representation of the derived age-metallicity relation for the bar field;
squares give the metallicity of the population with the highest star
formation rate at each age bin, while crosses give the mean 
star-formation weighted metallicity from the six metallicity bins. The
solid line shows the relationship from Pagel and Tautvaisiene (1998),
which does a reasonable job of matching observations of LMC clusters
(e.g., Olszewski 1993). Our derived relation is similar to the
model which was designed to fit the cluster observations; the width of
our relation at fixed age is also qualitatively similar to that seen in
the cluster distribution.

While the fits are significantly better using a range of metallicities
at a given age, the derived star formation histories are qualitatively
similar to those derived using the constrained dispersion-free
age-metallicity relation. As with the constrained age-metallicity fits,
we find evidence that the bar field contains a larger relative number
of older stars than the outer fields.  We find it encouraging that the
results on the star formation appear to be reasonably robust against
assumptions made about the metallicity distribution.

Since our models allow for multiple combinations of age and metallicity, the
results actually make a crude prediction for the metallicity distribution
of LMC stars; the prediction is crude since we are only using six discrete
metallicities in our models rather than a continuous distribution. 
Figures \ref{fig-hs2750_54} and \ref{fig-outer0_57} 
suggest that the LMC has a relatively broad metallicity
distribution. Low metallicity stars ([Fe/H] $\simlt$ -1) comprise
15 and 30\% of the stars in the outer and bar fields, respectively.
However, one needs to beware of directly comparing these numbers to 
observations of giant star metallicities. Our models give the relative
numbers of stars of all stellar masses at different metallicities, 
while giants sample only a small range of stellar masses. Because older,
more metal-poor, populations feed stars to the giant branch slower than young
populations, the relative number of lower metallicity \textit{giants} will be
lower than the true relative fraction of lower metallicity stars which
is predicted by the models. This effect can be substantial; for the bar model, 
we estimate that that low metallicity giants will be undersampled by
nearly a factor of two in a pure giant sample as compared with the true
metallicity distribution. Thus relatively few metal-poor giants are
predicted by these models.

Our models do have a reasonable component of relatively metal-rich
(solar or greater) stars, which are included to fit the reddest
sections of the main sequence. It is possible that the contribution of
these stars is overestimated by our models because of some contribution
from unresolved binaries, as discussed next.

\subsubsection{Unresolved binaries}

Unresolved binary stars can significantly affect the distribution of
objects in a color-magnitude diagram.  Although a variety of evidence
suggests that a significant fraction of all stars are in binary
systems, it is less clear whether the masses of stars in such systems
are correlated or are drawn from the same initial mass function. If
stellar masses of binaries are uncorrelated and the mass function rises
towards lower masses, then the effect of unresolved binaries is mainly
significant only for rather low mass stars at the bottom of the main
sequence; for more massive stars, a binary companion is much more
likely to be significantly fainter and thus have little influence on
the total system luminosity and color. As a result, the only way to get
a broadening of the main sequence for stars similar to those observed
by us in the LMC ($\sim$ 1\Msun) is to require that the masses of the
components of binary systems are correlated. Such a scenario has been
suggested by Gallart et al. (1999) to explain the color-magnitude
diagram of the Leo I dwarf spheroidal; they find significantly better
fits using a large fraction of binary stars which are constrained to
have mass ratios greater than 0.6. However, we find that models using
such a scenario still do not accurately reproduce our observed
color-magnitude diagram using a dispersion-free age-metallicity
relation.  The problem arises because the width of the observed
sequence is broadest compared to the models at intermediate
luminosities. If there is any range of mass ratios in binaries at all,
the effect of binaries grows with decreasing luminosity. Thus, any
model which matches the width of the main sequence at intermediate
luminosities using a binary component predicts too broad of a sequence
at lower luminosities.

To further check the binary star hypotheses, we performed fits with
multiple combinations of age and metallicity, but using an assumed
binary fraction of 0.5; binary masses were drawn from the same initial
mass function as the parent population but with binary mass ratios
constrained to be larger than 0.5. These fits are significantly worse
than fits without binaries. However, as mentioned above, if all mass
ratios were allowed, then the models would allow for a significant
component of binaries since their effect is small for the stars in our
observations.

Although we find that metallicity spread is a more likely explanation
than unresolved binaries for the observed width of the main sequence,
it is likely that both effects play some role. The existence of some
unresolved binaries would probably reduce some of the spread in our
derived age-metallicity relation; in particular, we expect it would
reduce the contribution of stars at the highest metallicities at any
given age.

\subsection{Accuracy of derived star formation histories from model fits}

Before one reads too much into these derived star formation histories,
however, one should consider some of the limitations and problems of
fitting star formation histories.

There are numerous assumptions that are made in the models:
\begin{itemize}
\item The stellar models accurately predict the observed properties of
stars as a function of age and metallicity,
\item A unique initial mass function exists which is independent of
age and metallicity, and is represented (in this case) by a power
law with $dN/dM \propto M^{-2.35}$,
\item All stars are found at a common distance and extinction,
\item The observational data are calibrated to the same system as the
models, and the observational errors can be accurately measured,
\item The basis stellar populations used in the fits represent all
populations present in the galaxy.
\end{itemize}

All of these assumptions are likely to be in error at some level. 
Consequently, the question is the degree to which deviations from the
assumptions affect the derived star formation histories. Unfortunately,
this is very difficult to assess given the unknown nature of the
possible errors in the assumptions.

As a result of these problems, it is likely that no solution will actually
matches the observed distribution of stars within the errors expected
from Poisson statistics alone. This is certainly the case for the best
models here; given the number of independent regions in the color-magnitude
diagrams being fit, one would expect a reduced $\chi^2$ much closer to
unity than the values we obtain.  Sometimes the fits produce model luminosity
functions which are consistent with the data, but other times they
are formally ruled out with a KS test.  Review of the various papers which 
derive star formation histories for various systems suggest the same
quality of matches is obtained for most other derived star formation
histories. Given the known problems with the assumptions, this is
usually not considered to be a major problem; instead, one makes the
assumption that the ``best-fitting'' model represents the closest
approximation to the truth, even if it is statistically inconsistent
with the data. We make the same assumption, but feel the need to
explicitly state it; one could certainly imagine situations in which
this assumption might not be true.

Because one is just choosing the best-fitting model averaged over the
entire color-magnitude diagram, our method inherently weights areas
where there are more stars and where the photometric errors are low.
As a result the model does not give extra weight to regions which
carry more unique information about stellar ages. For example, if upper
main sequence stars exist, there must be a young population, but if these
stars are vastly outnumbered by older stars, the model will do its best
to fit the older stars even if it means sacrificing a good match to the
younger stars. One could certainly devise a scheme where certain regions
of the color-magnitude diagram carry extra weight, and perhaps this
is the direction we should take in the future. 

In addition, different assumptions can lead to systematic errors in
the derived star formation histories. For example, we found that
changing the assumed reddening led to systematic differences in 
the derived star formation history. At a higher reddening, more metal-poor
stars are required to fit the observed data. If the age-metallicity
relation is constrained, then this in turn leads to a higher derived
number of older stars. Errors in the assumed initial mass function can
lead to similar systematic effects. 

\begin{figure}
\caption{Values of $\chi^2$ for best-fitting models for different values
of reddening (left) and distance modulus (right). Top panels are for bar
field, while bottom panels are for outer fields. The two different lines
in each panel correspond to fits with a fixed age-metallicity relation
and those will multiple combinations of age and metallicity.}
\plotone{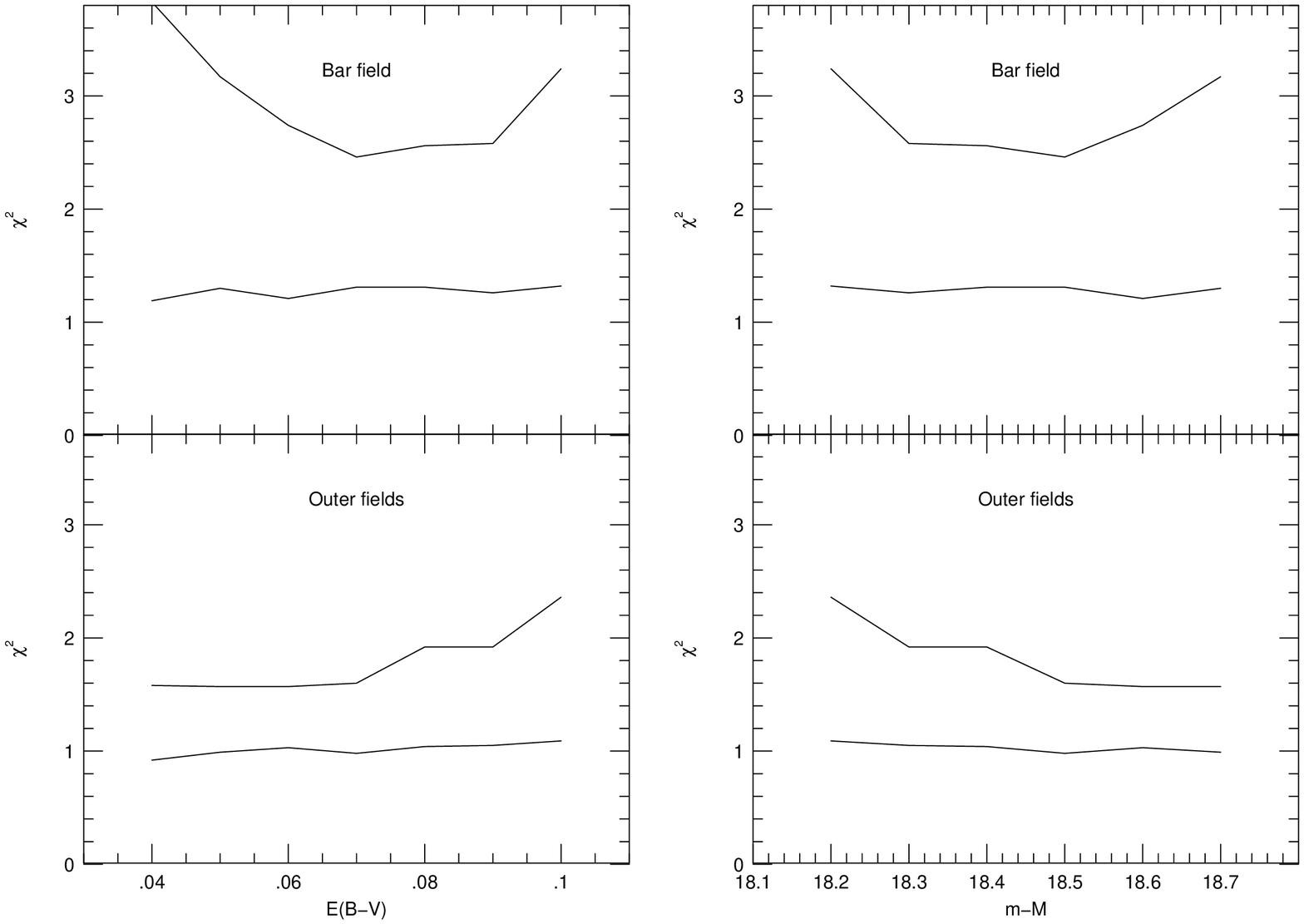}
\label{fig-chi2}
\end{figure}

To demonstrate the effect of varying reddenings, distances, and IMF slopes,
we ran a set of solutions allowing for a range in reddening from
$0.04<E(B-V)<0.10$, a range of distance from $18.2<m-M<18.7$, and two
different IMF slopes with $\alpha=-2.35$ and $\alpha=-2.95$; each of
these was varied independently with the other two quantities at our
preferred values. For each different parameter, we derived a star
formation history along with a $\chi^2$ for the each fit. Figure \ref{fig-chi2}
shows the derived values of $\chi^2$ for different choices of reddening
and distance modulus; in each panel, results are shown both for the
constrained age-metallicity relation as well as for multiple combinations
of age and metallicity. If the age-metallicity relation
is constrained, the quality of the fit changes significantly for different
choices of reddening and distance modulus; minimum $\chi^2$ are reached
around our preferred values of $m-M=18.5$ and $E(B-V)=0.07$ and 0.04 for the
bar and outer fields. However, if multiple combinations of age and metallicity
are allowed, the quality of the fits are relatively insensitive to the
choice of reddening and distance modulus, indicating there is some
degeneracy in the sensitivity to different parameters. 

This supports our assertion that it is better to use additional
independent observational constraints on parameters relevant to the star
formation fit rather than to include these parameters in the fitting
process. For many systems, information about distance and reddening
are readily available. We suggest that perhaps the greatest improvement
in the confidence in our derived star formation histories will come
from the observations of observed metallicity distributions against
which one could compare the derived star formation histories. The
derivation of metallicities is feasible in nearby stellar systems given
current multi-object spectroscopic capabilities and/or using multi-band
photometry, and, in fact, such studies are underway by several groups
(e.g., Olszewski, Suntzeff, \& Mateo 1996, Smecker-Hane et al., private
communication). However, we reiterate that caution must be used to
compare observations with model predictions; one must take into account
the metallicity distribution biases which are introduced by the
observational selection of stars used for metallicity determinations.

\begin{figure}
\caption{Derived star formation histories for a variety of choices of
$m-M$, $E(B-V)$, and IMF slope, as described in the text. Top panels are
for outer fields and the bottom panels are for bar field. Left panels
show results with a constrained age-metallicity relation, while right
panels allow for multiple combinations of age and metallicity. Bold line
indicates results for our preferred set of parameters; dashed line shows
the results for a steeper IMF.}
\plotone{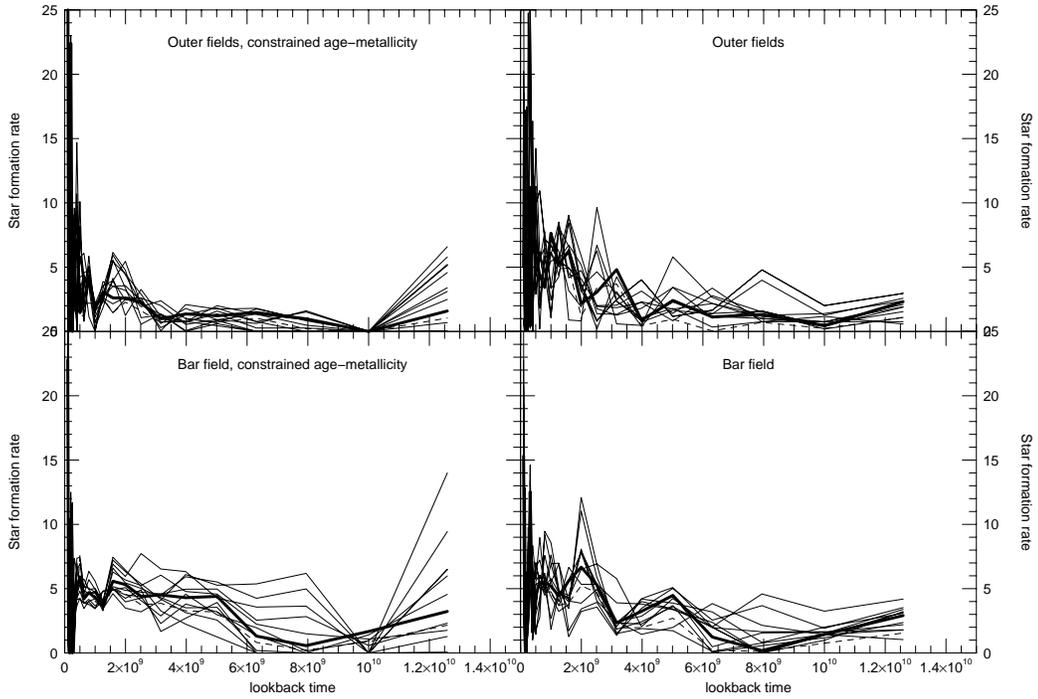}
\label{fig-sfh}
\end{figure}

Given the limited external constraints we have about the distance modulus 
to the LMC, the extinction, the IMF, and the metallicity distributions,
we must consider how possible uncertainties in these affect our derived
star formation histories. Figure \ref{fig-sfh} shows derived star formation
histories for the ensemble of models comprising $18.2<(m-M)<18.7$, 
$0.04<E(B-V)<0.10$, and IMF slopes of -2.35 and -2.95; the bold line shows
the results from previous figures for our preferred quantities. One can see that
the star formation history is qualitatively similar independent of
the parameters, but quantitatively, the star formation rate at any
given time can be in error by a factor of a few. The largest qualitative
difference comes for different choices of IMF slope; as the IMF becomes
steeper (dashed line in Figure \ref{fig-sfh}), 
the observations require a larger relative number of younger stars,
exactly as expected (Holtzman et al. 1997).  To the extent to which 
our parameter choices span the full range of values expected for the
LMC, Figure \ref{fig-sfh} can be used to give a reasonable estimate of
the uncertainties in our results, although these results do not consider
the possible effect of errors in the stellar models.

In addition to the systematic errors, there are random errors in our
results because of the limited number of stars observed. These random
errors are larger for the outer fields than for the bar field because
they have fewer observed stars. However, simulations of color-magnitude
diagrams suggest that the magnitude of the random errors, even for the
more sparse outer fields, are smaller than those which arise because
of potential systematic errors.

\subsection{Differential comparison of color-magnitude diagrams}

To avoid potential problems with fitting star formation histories, it is
possible to derive \textit{differences} in the star formation history
from one field to another by a direct comparison of the observed
color-magnitude diagrams. Such differences lead to systematic residuals
when comparing the Hess diagrams of different fields. Differential
comparisons of fields with similar metallicities are relatively
straightforward to interpret in terms of age differences, although a
quantitative association of a difference with an age requires the use
of stellar models.  Differential comparisons may be more problematic for
fields with significantly different metallicities.

As an application, we consider the differences in star formation history
between the LMC outer fields and the bar field. Our model-dependent derived 
star formation history suggested that there has been a greater relative
contribution of the youngest stars in the outer fields than in the bar,
in agreement with the results derived by Olsen (1999) based on similar
fits, but in qualitative disagreement with the results of Elson et al. (1997)
which were based on a visual inspection of the color magnitude diagram
in a bar field.

\begin{figure}
\caption{Difference of the Hess diagrams between the bar and the outer fields,
smoothed to reduce noise from counting statistics.}
\plotone{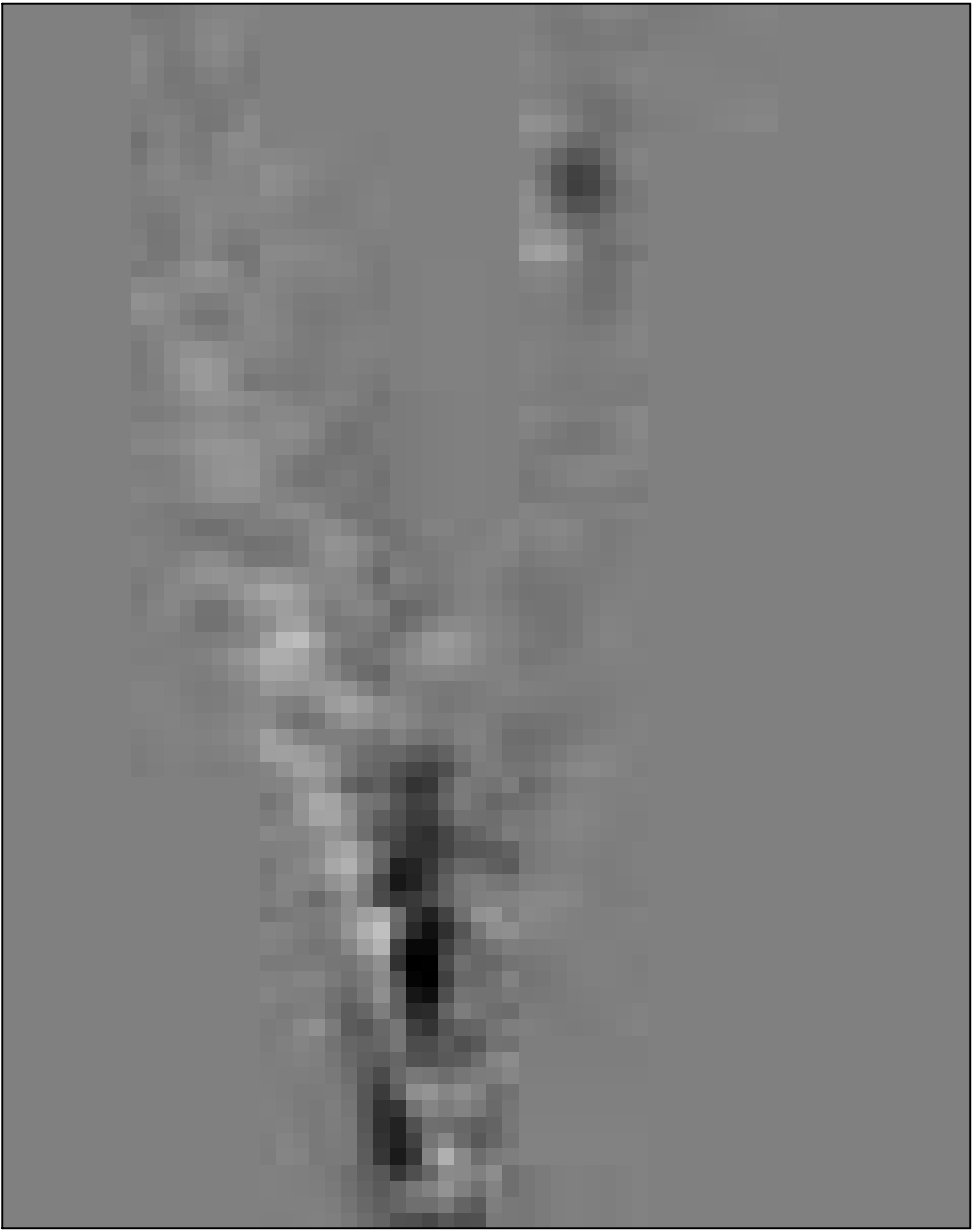}
\label{fig-outerbarcomp}
\end{figure}

Figure \ref{fig-outerbarcomp} shows the difference in the Hess diagrams
between the two fields, where white areas represent locations where
there are more outer field stars, and darker areas regions where there
are more bar stars. The Hess diagrams were normalized to have the same
total number of stars between $M_V\sim 4-4.5$ where the photometry in
both fields is reasonably complete and where stellar evolution effects
are minimal; the difference between the Hess diagrams was smoothed to
suppress the noise from counting statistics.  One can clearly see a
darker band in the lower parts of the residual Hess diagram which
suggests that the bar contains a relatively larger number of
intermediate age stars than the outer field; the difference is made up
by a relatively larger number of upper main sequence (younger) stars
in the outer fields.  The bar field also has a relatively larger number
of red clump stars which represent stars of intermediate age.
Consequently, the differential, model-independent comparison supports
the results derived by fitting stellar models, namely that the bar,
although it contains a significant population of young stars, is
\textit{relatively} older than the outer fields.

\begin{figure}
\caption{Synthetic Hess diagrams for a continuous star formation rate over
12 Gyr with our adopted age-metallicity relation for the LMC.}
\plotone{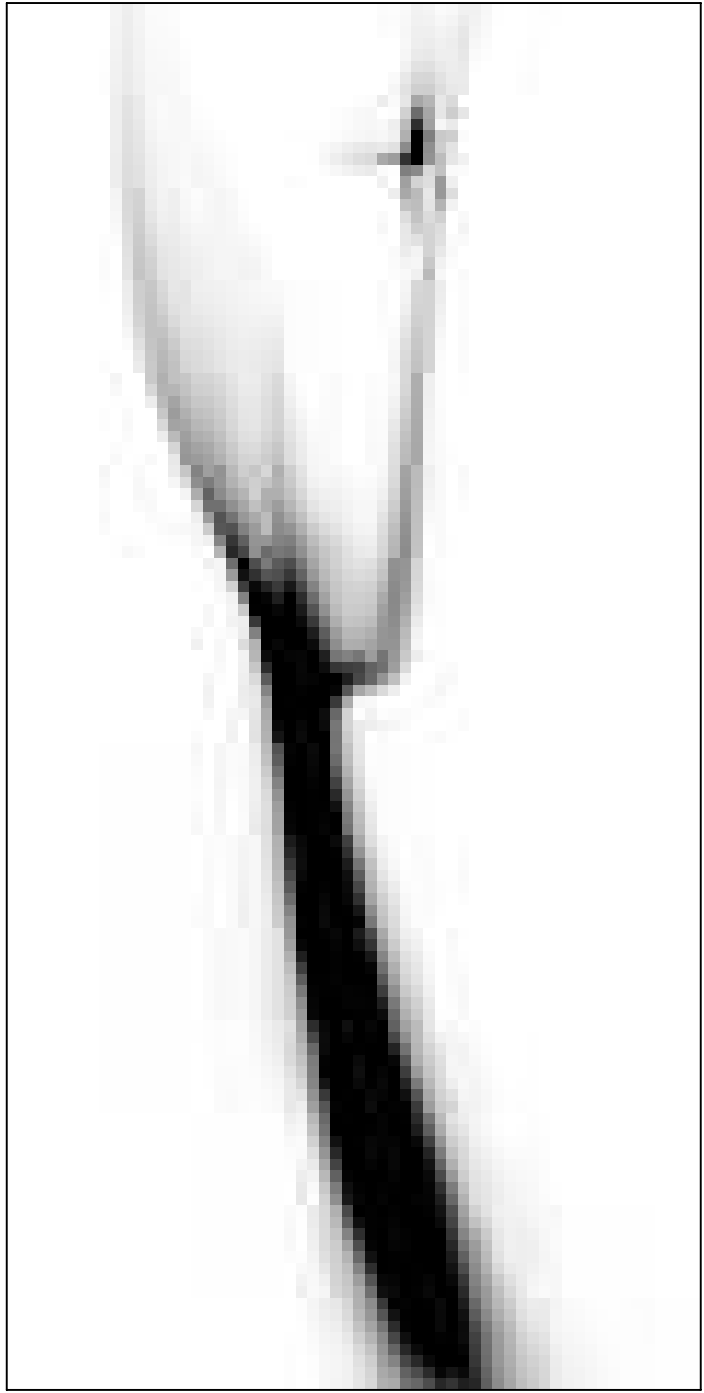}
\label{fig-continuoussim}
\end{figure}

Although the bar field shows an apparently significant sequence which
one might associate with a several Gyr old burst (as seen in Figure
\ref{fig-cmd} and shown in cross-section plots in Elson et al. 1997),
which shows up in contrast to the outer field color-magnitude diagrams,
such a feature turns out to be a generic feature of models even with a
\textit{continuous} star formation rate. This arises because upper main
sequence stars (those with convective cores) evolve to cooler
temperatures and higher luminosities over most of their main-sequence
lifetimes, but then retreat to higher temperatures, creating a jag in
the evolutionary path in a color-magnitude diagram.  Since the star
spends proportionally more time at the coolest effective temperature, a
secondary sequence which is offset from the main sequence exists for a
continuous star formation history. Although this is true even for a
population of fixed metallicity, an age-metallicity relation makes the
secondary sequence even more pronounced.  The effect is demonstrated in
Figure \ref{fig-continuoussim}, where we show a synthetic Hess diagram
of a population with a constant star formation rate over the past 12
Gyr, using our adopted age-metallicity relation for the LMC.  One sees
a clear sequence which might be confused with an increase in the star
formation rate at some time in the past, despite the fact that it is a
color-magnitude diagram for a constant star formation rate.  This
clearly shows the peril of interpreting color-magnitude diagrams purely
visually; an apparent concentration of points does not necessarily
imply a burst or even a significant enhancement in the star formation
rate.

This observation leads to an understanding of the difference between
the interpretation of Elson et al. (1997) and the results of this paper
and Olsen (1999) regarding the relative age of the LMC bar. Elson et
al. (1997) suggest that the LMC bar is younger than the rest of the LMC
because they observe a bimodal distribution of color in the upper main
sequence of their LMC bar field. They associate the blue peak with the
formation of the LMC bar ($\sim$ 1 Gyr ago) and the red peak with the
formation of the bulk of LMC field stars ($\sim$ 4 Gyr ago). Instead,
we find that the red peak is a generic prediction of the models even
for roughly constant star formation rate, and the blue peak represents
a recent increase in the star formation rate which is seen in both the
outer fields and the bar (in fact, it is stronger in the outer fields
than in the bar).

\subsection{Field vs. cluster star formation history}

Perhaps the most notable conclusion which can be drawn from our derived
star formation histories is that the field star formation history in
both the bar and the outer fields appears to differ from the star
formation as suggested by the age distribution of LMC clusters. LMC
clusters show a significant age gap between lookback times of 4 and 12
Gyr (e.g., van den Bergh 1991, Girardi et al. 1995), with 14 old
clusters (which have ages comparable to those of the Galactic
globulars, see Olsen et al. 1998), numerous young clusters, and only
one cluster, ESO 121-SC03, at an intermediate age. In contrast, the
derived star formation histories of Figures
\ref{fig-hs2750_44}-\ref{fig-outer0_57} suggest that star formation has
been more continuous in the field of the LMC. Here we consider the
robustness of that conclusion.

Geha et al. (1998) showed that the observed luminosity function in
the outer fields was strongly inconsistent with a star formation
history which corresponds to the current \textit{number} distribution of 
clusters as a function of lookback time. However, this comparison is
perhaps unfair, as the older clusters are generally more massive than
the younger ones, so weighting by mass would allow for a larger older
component. In addition, one might consider that some fraction of
clusters which were formed at an early epoch might be disrupted during
the subsequent evolution of the LMC; although many of the young
clusters are massive and appear tightly bound and unlikely to disrupt
anytime soon, many others have lower masses and larger sizes and might
plausibly disrupt.

\begin{figure}
\caption{Derived star formation history for bar field, assuming an 
age-metallicity relation for the LMC, but constraining the fit to have no
star formation between 4 and 10 Gyr ago.}
\plotone{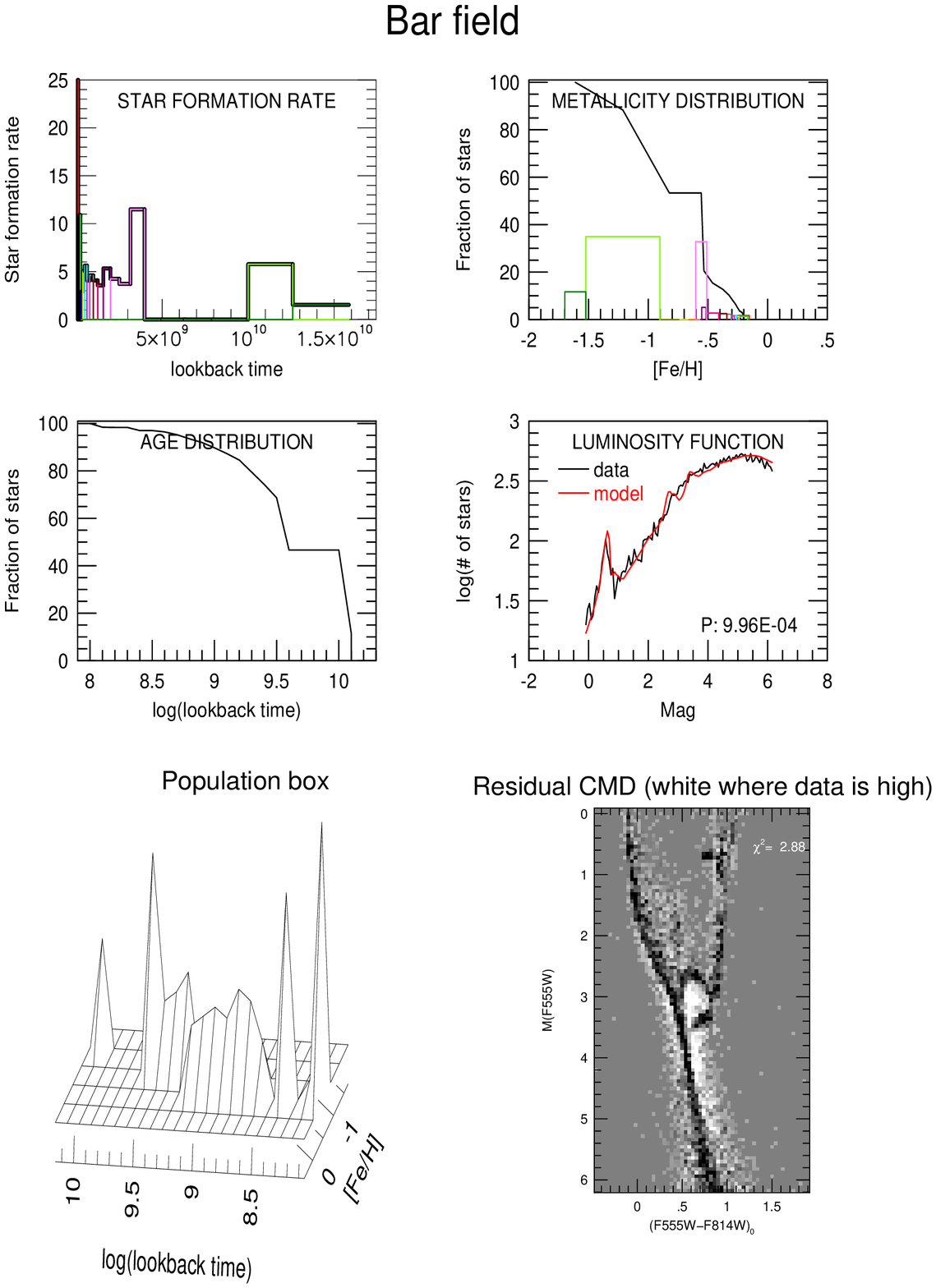}
\label{fig-hs2750_64}
\end{figure}

\begin{figure}
\caption{Derived star formation history for bar field, allowing for multiple
combinations of age and metallicity, but constraining the fit to have no
star formation between 4 and 10 Gyr ago.}
\plotone{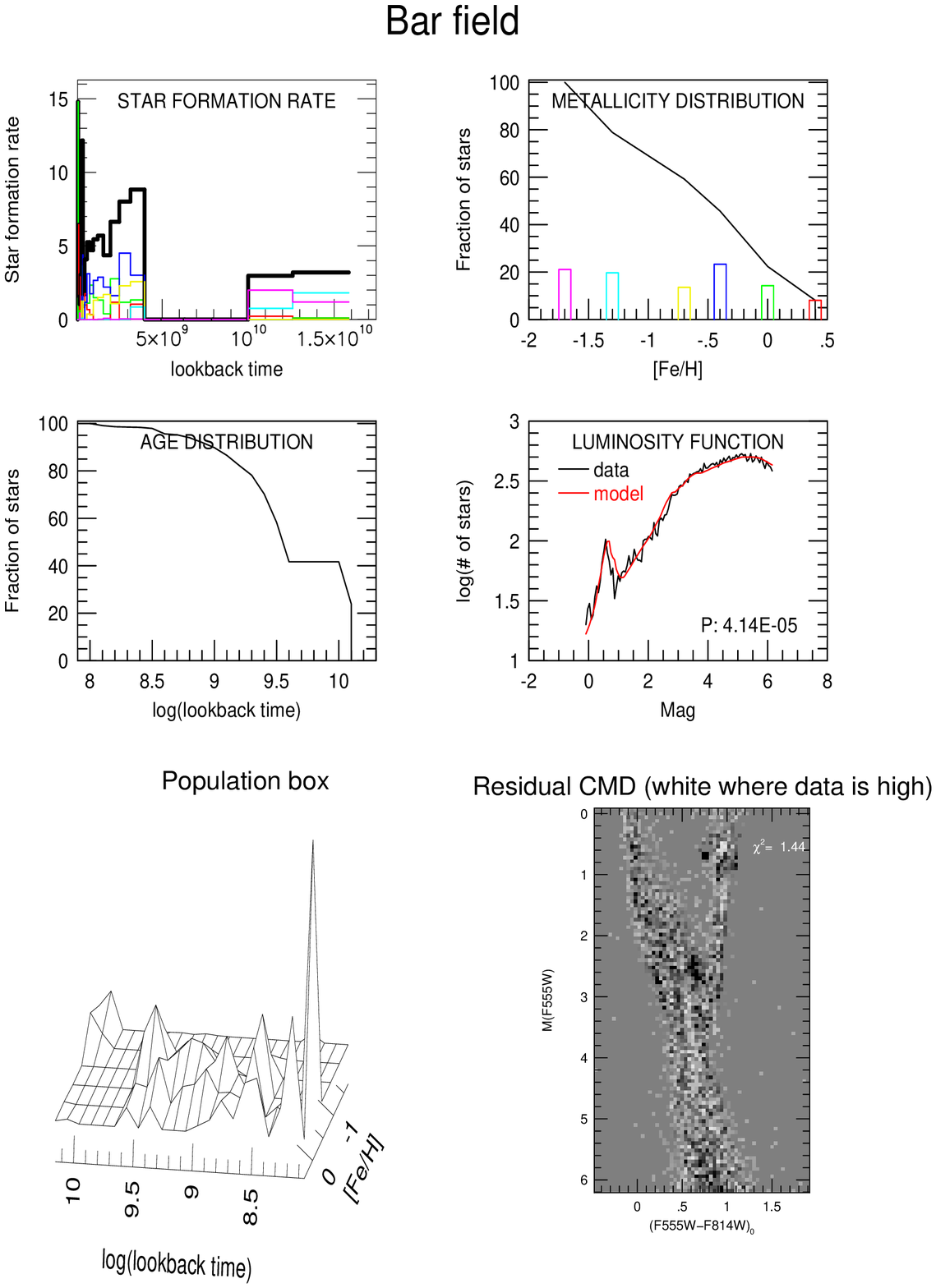}
\label{fig-hs2750_74}
\end{figure}

As a result, we consider the more general question of whether any star
formation history with a gap in star formation between 4 and 10 Gyr is
capable of reproducing the observed properties of the LMC field stars.
To address this, we performed fits for the star formation history again
without allowing for \textit{any} component stellar populations with
ages between 4 and 10 Gyr. Figures \ref{fig-hs2750_64} and
\ref{fig-hs2750_74} show the results for the bar field for the
constrained age-metallicity relation and arbitrary combinations of age
and metallicity. The fit with the constrained age-metallicity relation
is notably worse than allowing for intermediate age stars. This is
easily explained; the existence of a broad band of subgiants around the
oldest turnoff suggests that multiple ages are present. However, it is
possible to get such a continuous band with different combinations of
age and metallicity, since older, lower metallicity stars can blend
smoothly into younger, higher metallicity stars without necessarily
leaving a gap in the color-magnitude diagram. This is confirmed by
Figure \ref{fig-hs2750_74}, which shows that a moderately good match to
the observed Hess diagram can be made even with an age gap in the star
formation history. However, one can see that the model produces too
many subgiants at $M(F555W)\sim 2.5$; this can be seen in both the
residual Hess diagrams as well as in the luminosity function. The
$\chi^2$ for the star formation history is only slightly worse with an
age gap than without it, but the probability that the luminosity
function is consistent with that of the data can be ruled out at a much
higher confidence level than for models without a gap. In addition, the
existence of a gap would require a relatively larger population of
older, metal-poor stars; with a gap, we find that approximately 40\% of
all of the LMC field stars would have to be older than 10 Gyr and more
metal-poor than [Fe/H] $\sim$ -1.  This may not be consistent with
observations of metallicity distributions (e.g.  Olszewski 1993) and
the lack of a strong horizontal branch in the color-magnitude diagrams;
however, the possibility exists that the LMC has an extended, low
density, older stellar halo which becomes more dominant over the young
and intermediate age population as one moves farther from the center of
the LMC.

Consequently, we feel that it is unlikely that the field star formation
history, as sampled by the location of our fields, has a gap in star
formation between 4 and 10 Gyr ago. 

%It is thus possible that a global average star formation
%history could more closely match the cluster age distribution. Still,
%different mechanisms for driving star formation in clusters and in
%field stars would be suggested, since the young and old cluster systems
%do not appear to have significantly different spatial distributions (is
%this true?).

\subsection{Burstiness of star formation}

Another outstanding question is the degree to which star formation is
``bursty'' in the LMC. The degree to which we can distinguish between
bursty and continuous star formation depends on several factors.
For older populations, the distribution of stars in the color-magnitude
diagrams changes very slowly with age, so it is difficult to get
much age resolution. For younger populations, the separation between
ages is larger, but the observed number of stars is smaller, so
sensitivity to different distributions of star formation is limited
by counting statistics.

Our fits for star formation history have been performed assuming
constant star formation within epochs spaced by 0.1 in $\log(age)$.
This value for the width of each epoch was determined by finding
the narrowest age bin which gave statistically distinguishable fits, as
measured by $\chi^2$. 

To measure the sensitivity of the technique to burstiness in the star
formation rate, we performed the star formation fits in which single
age bins were given a duration of $\Delta(\log(age)) = 0.01$, while
preserving the 0.1 spacing in $\log(age)$. We did this for each age
bin in turn for lookback times from 1 to 4 Gyr. We found that we
obtained nearly identical quality fits using the 0.01 width epochs as
we did with the 0.1 width epochs, although the fits would have been
worse if we had required more than one bin to be "bursty" at a time. 
The basic reason we could not discriminate the duration of a star
formation epoch is small number statistics in the number of stars
observed on the upper main sequence; without the counting
statistics, the models are straightforward to distinguish.  This was
true even for the bar field, which is the densest field one could
observe in the LMC. We estimate that increasing the number of stars by a factor
of 5-10 would allow burstiness to be distinguished, suggesting that
a program with multiple pointings with WFPC2 (e.g., Smecker-Hane et al.,
in progress) and/or the Advanced Camera would be useful. At ages older
than 4 Gyr, burstiness becomes extremely difficult to measure without
exquisitely accurate photometry.

\subsection{Star formation and the interaction history of the LMC}

It has been suggested that star formation in the LMC is triggered by
tidal interactions with the Milky Way and the SMC. As a result, it is
of interest to see whether there is any evidence for an enhanced star
formation rate around the time of the last closest passage. Since the
full orbits of the Magellanic Clouds are still unknown, the time of
last close passage is somewhat uncertain, but the latest models place it
around 2.5 Gyr ago (Zhao, private communication). Inspection of our
derived star formation histories (for example, Figure \ref{fig-sfh}) 
show a general tendency for the star formation to increase by a mild
amount around this time, but no dramatic effect is seen.
However, as discussed in the last section, we cannot constrain
the burstiness of the star formation rate very accurately from the
current data, so some correlation of star formation with orbit is not
necessarily ruled out.

One might expect that triggered star formation would not
occur in all regions of the galaxy at the same time. If star formation
were triggered in different regions at different times, subsequent
mixing arising from the stellar velocity dispersion and differential
rotation would smooth the bursty nature of the triggered star formation
on a timescale given by the mixing. Given an approximate velocity dispersion
of 50 km/s, it would take only $\sim$ 500 Myr for stars at a radius
of 4 kpc to mix azimuthally.

\subsection{Chemical evolution and star formation history of the LMC}

Pagel and Tautvaisiene (1998) have recently published a model for the
chemical evolution of the LMC and compared it with previous
models. Such models attempt to match the observed abundance distributions
of different elements as a function of metallicity. In general, these
models allow to star formation rate to be a free parameter. Models
differ in the adopted yields, IMF, the presence of inflow and/or outflow, and
the degree to which outflow is selectively enhanced in heavy elements.

The best fitting model of Pagel \& Tautvaisiene for the LMC favors a
star formation history which, although they call it a ``bursty'' model,
has an underlying constant star formation rate over the history of
the LMC, with an enhancement in the star formation rate 3 Gyr ago. 
However, they also have a model with a smoothly varying star formation rate
which also provides a reasonable match with observational data. We
suspect that using a star formation history derived from our color-magnitude
diagrams would be able to provide a reasonable match to the abundance
data as well, as it is intermediate between the two models presented
in Pagel and Tautvaisiene. 

We suggest that the next logical step in modeling star formation histories
is to couple the derived star formation rates with chemical evolution
models and simultaneously attempt to match both color-magnitude diagram
data and abundance data. In principle, this might allow one to more uniquely
determine the importance of mass inflow/outflow. Unfortunately, such 
attempts will be complicated if the star formation history is a strong
function of location in the galaxy. However, from the few fields considered
to date, it appears that the variation with position may not be so large
as to make such an attempt futile; once data on more fields, particularly
at large radii become available, we suspect a simple model with a small
number of radial zones might be sufficiently accurate.

\section{CONCLUSIONS}

We have derived star formation histories from the distribution of stars
in deep color-magnitude diagrams obtained using HST. These data suggest
that there is a significant component of stars older than 4 Gyr in both
outer fields and the bar of the LMC. Models in which there is a dispersion-free
age-metallicity relation cannot reproduce the width of the main sequence
in our high accuracy photometric data. As a result, we have fit models
which allow for multiple combinations of age and metallicity and find we
can obtain accurate matches to the observed data. These fits allow us to
fully construct ``population boxes'' from our data which are derived solely
from color-magnitude diagrams. Such diagrams qualitatively reproduce the
mean age-metallicity relation observed in LMC clusters as well as the
spread around this relation. These derived models produce crude predictions
for the abundance distribution the LMC; new observations which provide
such distributions will be extremely useful in constraining the star
formation histories and confirming the validity of the models. 

Both the model fits as well as a differential comparison between the 
observed color-magnitude diagrams suggest that the bar of the LMC contains
a relatively larger number of older stars than the outer fields. This
is consistent with the conclusions of Olsen (1999) but different from
those of Elson et al. (1998), although we have presented 
a plausible explanation for why the latter study reached their conclusions.

One main implication of the derived star formation histories is that
the field star formation history appears to differ from that suggested by
the LMC clusters, in that there does not appear to be an age gap in the
field star age distribution. However, we note that it is actually rather
difficult to constrain the star formation history for lookback times greater
than 4 Gyr given the age-metallicity degeneracy in the location of isochrones;
observations of larger samples of subgiants, ideally with metallicity
determinations, would be desirable to confirm that field stars fill
the cluster age gap.

We find that it is quite difficult to constrain the degree to which star
formation is bursty in the LMC on time scales less than about 25\% in
age with the observed number of stars even in the WFPC2 bar field. Larger
samples will be required to address this issue. However, sequential star
formation across the LMC followed by mixing may erase the signatures of
bursty star formation even if it occurs.

Future progress will be made with larger samples of stars; with
accurate photometry down to the oldest main sequence turnoff, one can
further constrain burstiness and the star formation history. In
addition, we suggest that metallicity determinations for a large number
of stars will be crucial in constraining and testing derivations of
star formation histories. Coupled with chemical evolution models, we
may be able to get constraints on the importance of inflow/outflow in
the LMC, and begin to fully understand the nature of the star
formation history in one of our nearest neighbors.

This work was supported in part by NASA under contract NAS7-918 to JPL.

\newpage

\begin{center}
\begin{tabular}{c c c c c c}
\multicolumn{5}{c}{\bf Table 1.} \\
\multicolumn{5}{c}{ Summary of Observations} \\ \hline \hline
Field & $\alpha_{2000}$ & $\delta_{2000}$ & Exposure time & STScI filenames \\
\hline
LMC-OUTER  & $5^h 14^m 44^s$ & $-65^{\circ} 17' 43''$ & 4000s & u2c5010[1-8]t  \\
HODGE-10   & 5 58 21 & -68 21 19 & 2500s  & u2o9020[1-6]t \\
HS-275 (BAR)&5 24 21 & -69 46 27 & 3700s & u2o9010[1-8]t 
\end{tabular}
\end{center}

\end{document}